\documentclass[pra,10pt,twocolumn,aps,floatfix,amsmath,amssymb,amsfonts]{revtex4-1}
\usepackage{dcolumn}
\usepackage{bm}
\usepackage{amsmath,amsthm,amssymb}
\usepackage{float}
\usepackage{graphicx}
\usepackage{subcaption}
\usepackage{color}

\usepackage{braket}

\def\beq{\begin{equation}}
\def\eeq{\end{equation}}

\begin{document}

\title{Quantum state engineering with a trapped atom and a set of static impurities}

\author{Marta Sroczy{\'n}ska$^1$, Tomasz Wasak$^1$, Krzysztof Jachymski$^{1,2}$, Tommaso Calarco$^3$ and Zbigniew Idziaszek$^{1}$}
\affiliation{
$^1$ Faculty of Physics, University of Warsaw, ul.~Pasteura 5, PL-02-093 Warsaw, Poland,\\
$^2$ Institute for Theoretical Physics III \& Center for Integrated Quantum Science and Technology (IQST), University of Stuttgart, Pfaffenwaldring 57, D-70550 Stuttgart, Germany\\
$^3$ Institute for Complex Quantum Systems \& Center for Integrated Quantum Science and Technology (IQST), Universit\"at Ulm, Albert-Einstein-Allee 11, D-89075 Ulm, Germany}
\date{\today}

\begin{abstract}
Hybrid systems of ultracold atoms and trapped ions or Rydberg atoms can be useful for quantum simulation purposes. By tuning the geometric arrangement of the impurities it is possible to mimic solid state and molecular systems. Here we study a single trapped atom interacting with a
set of arbitrarily arranged static impurities and show that the problem admits an analytical solution. We analyze in detail the case of two impurities, finding multiple
trap-induced resonances which can be used for entanglement generation. Our results serve as a building block for the studies of quantum dynamics of complex systems.
\end{abstract}

\maketitle

\section{Introduction}

Ultracold trapped atoms have found numerous applications in the field of quantum simulations of many-body physics~\cite{Bloch2008}. Properties of ultracold atomic systems
can be tuned in experiment using external electromagnetic fields, which provide the opportunity to shape the trapping potential experienced by the
atoms~\cite{Lewenstein2007} as well as their interactions~\cite{Chin2010}. Both bosonic and fermionic atomic species are available. These favorable properties lead to a
number of accomplishments with ultracold atoms in optical lattices such as observation of superfluid–-Mott insulator transition \cite{Greiner2002A}, superexchange
interactions for simulations of spin lattice Hamiltonians~\cite{Trotzky2008}, many-body localized phases of matter~\cite{Schreiber2015} or exotic quantum states such as
the supersolid phase \cite{leonard2017supersolid, li2017stripe}.  Quantum computation schemes involving cold atoms have also been proposed basing on various mechanisms
such as state-dependent potentials, exchange interactions, trap-induced resonances and
other~\cite{Jaksch2000,Stock2003,Charron2006,Hayes2007,Doerk2010,Negretti2011,Jachymski2014}.

In recent years, great progress has been made in realization of other quantum technology platforms such as trapped ions and Rydberg
atoms~\cite{Leibfried2003,Saffman2010,Wineland2013nobel,Haroche2013nobel}. Interestingly, trapped ions and cold atoms can be combined into a novel hybrid quantum
system~\cite{Tomza2017}. A chain of trapped ions can act as a periodic external potential for cold atoms, emulating a solid state with atoms playing the role of mobile
electrons~\cite{Bissbort2013}. Another promising hybrid system involves trapped Rydberg atoms acting as impurities instead of ions. Rydberg atoms can be
arranged in arbitrary three-dimensional structures using optical tweezers~\cite{Barredo2016,Endres2016,Bernien2017,Barredo2017}. In a similar way to the solid state simulation~\cite{Bissbort2013}, one can view hybrid systems as potential simulators of complex molecular phenomena such as formation and reconfiguration of chemical bonds or excitation transport in macromolecules. Here the atoms would play the role of electrons and the impurities would mimic nuclear cores.

To further increase the potential of such systems and uncover their novel applications, the interaction of a single atom with other particles needs to be understood
first. This is similar to finding natural orbitals of a molecular system. In this work, we make a first step in this direction by showing that the problem of finding the
eigenstates of a single harmonically trapped atom interacting with arbitrarily many impurities can be approached analytically in the limit of the low collision energy,
when the atom-ion interaction can be modeled with the $s$-wave regularized delta pseudopotential~\cite{Huang1957}. We provide a general method of solving the
Schr\"odinger equation describing such a system based on free function method~\cite{Antezza2010}, along with its application to a simple case of two impurities.

\begin{figure}[H]
  \begin{subfigure}[b]{\textwidth}
    \includegraphics[width=0.47\textwidth]{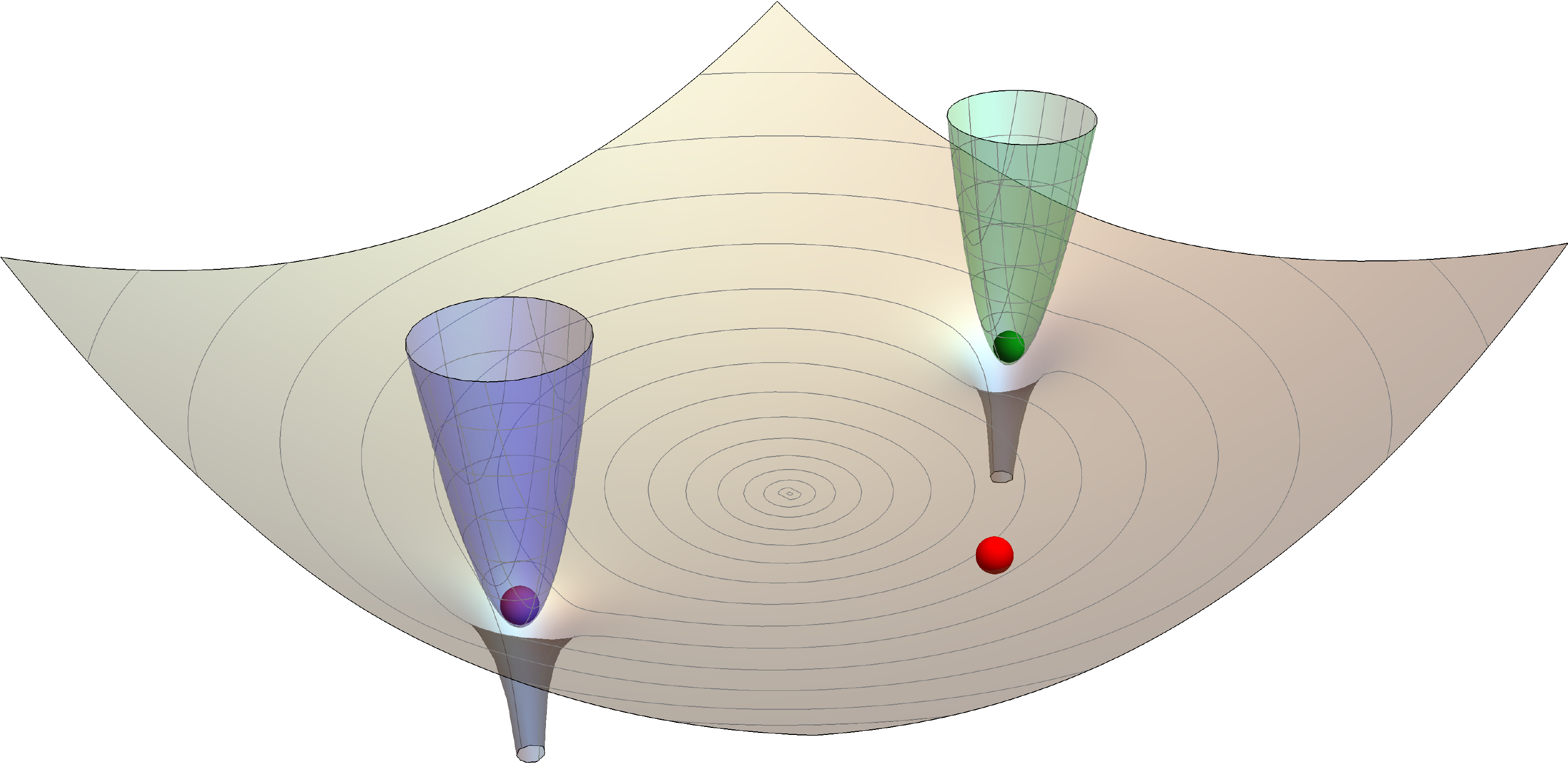}
  \end{subfigure}
  \caption{(Color online) An interaction potential experienced by a trapped atom (red sphere) in the presence of two localized impurities (purple and green spheres),
    which are localized by external trapping potentials (purple and green). In the vicinity of the static particles, the trapping potential
    is modified by the atom-impurity interaction; the gray surface is the  effective potential experienced by the atom.} \label{Fig:TIMechanism}
\end{figure}

\begin{figure*}
  \begin{subfigure}[b]{0.32\textwidth}
    \includegraphics[width=\textwidth]{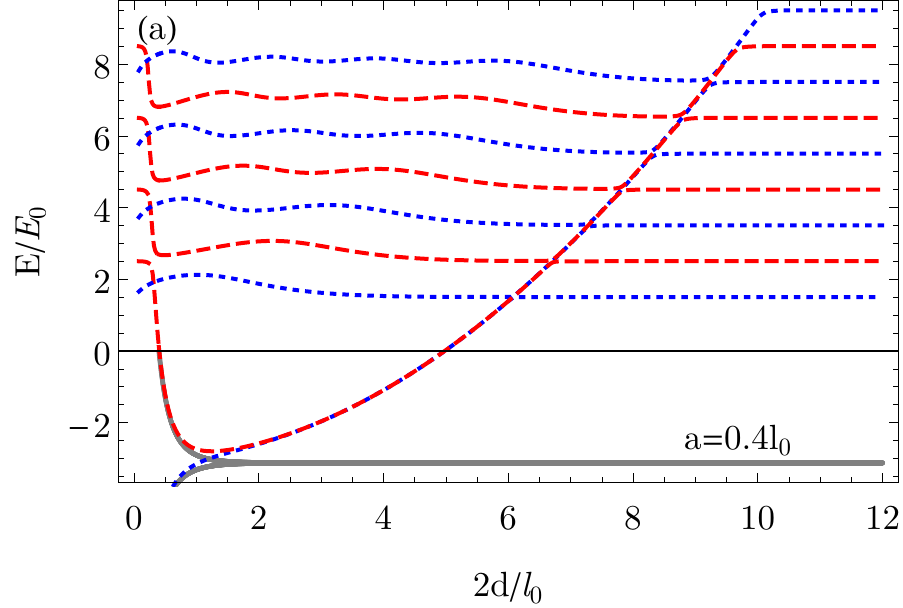}
  \end{subfigure}
  \begin{subfigure}[b]{0.32\textwidth}
    \includegraphics[width=\textwidth]{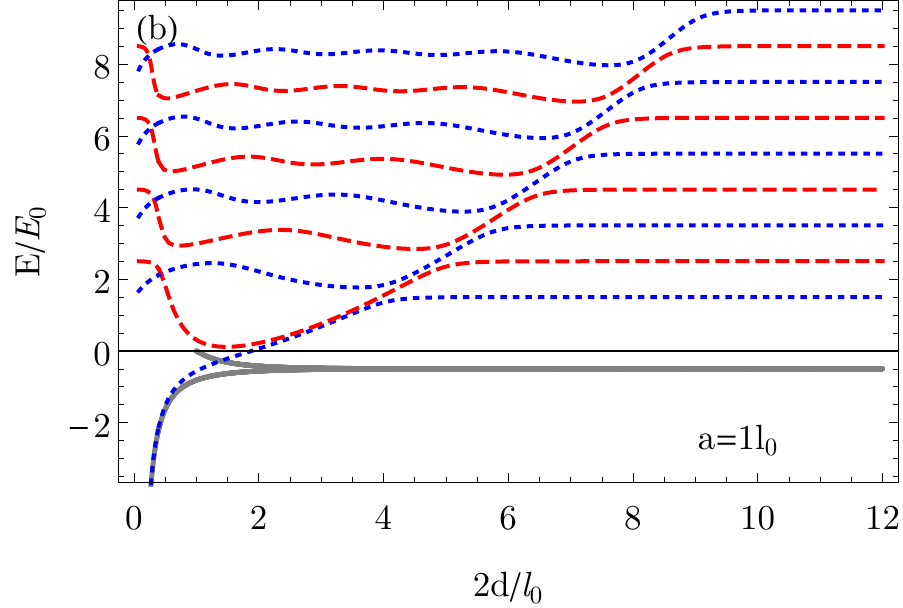}
  \end{subfigure}
  \begin{subfigure}[b]{0.32\textwidth}
    \includegraphics[width=\textwidth]{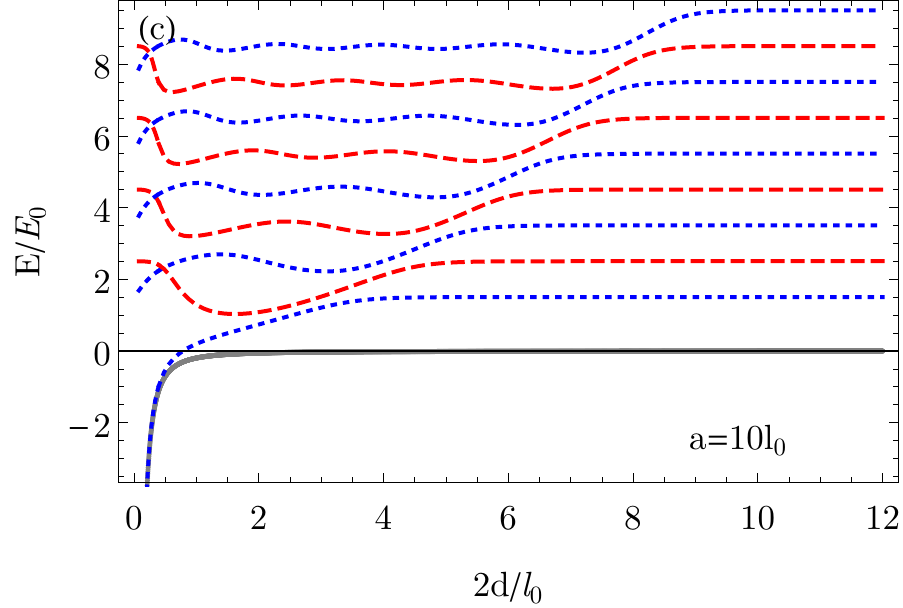}
  \end{subfigure}\\
  \begin{subfigure}[b]{0.32\textwidth}
    \includegraphics[width=\textwidth]{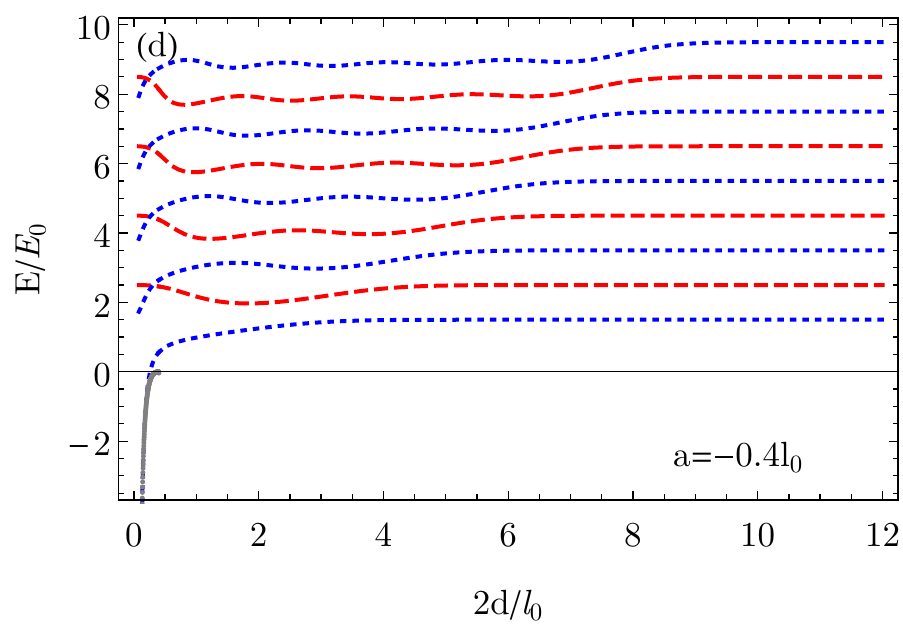}
  \end{subfigure}
  \begin{subfigure}[b]{0.32\textwidth}
    \includegraphics[width=\textwidth]{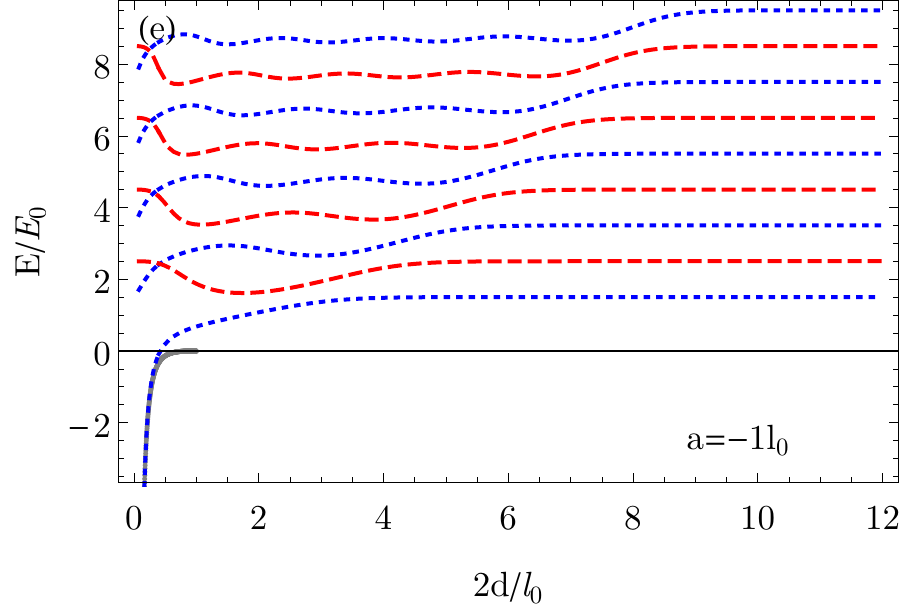}
  \end{subfigure}
  \begin{subfigure}[b]{0.32\textwidth}
    \includegraphics[width=\textwidth]{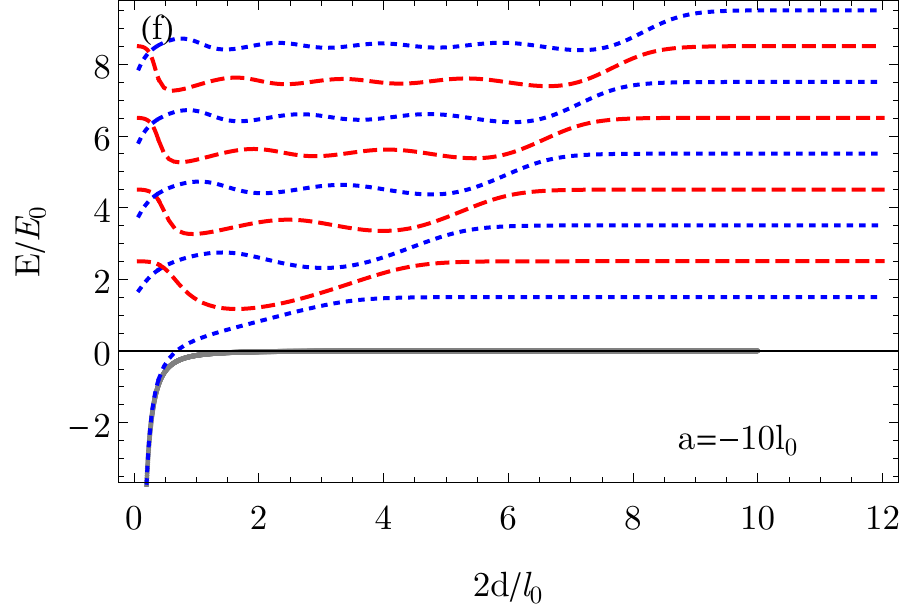}
  \end{subfigure}
  \caption{(Color online) The lowest energy levels of the atom as a function of the distance between the impurities, equal to $2d$, for different values of the scattering
    length $a$.  The dotted blue and dashed red lines denote even and odd states, respectively. The solid gray lines display the energy levels of the bound states ($E<0$)
    in absence of the trap.} \label{fig:energyLevels}
\end{figure*}


The structure of this manuscript is organized as follows. In section~\ref{sec:Model}, we first introduce the general Hamiltonian of a trapped atom interacting with many
static impurities. Then we present the method based on Green's function formalism that reduces the problem of solving the Schr\"odinger equation to a search of the roots
of a single function expressed in terms of the Green's functions. The details of the derivation are described in Appendix~\ref{app-sol}. In Section~\ref{sec:Results}, we
apply this general method to a system consisting of a single harmonically trapped atom interacting with two static impurities. For the case of symmetrically placed
impurities, we calculate the lowest energy levels as a function of the distance between the impurities and the atom-impurity scattering length.  In addition, we compare
the obtained results with a variational approach exploiting a simple trial wave function. We provide also the analysis of the avoided crossings that appear in such a
system due to the trap-induced resonances. 
Summary of the results and the feasibility of the molecular simulator are provided in Section~\ref{sec:Conc}.

\section{Model}
\label{sec:Model}

In this section, we first describe the Hamiltonian of the system, and then present the method of solving the stationary states of the system. The procedure we follow is
based on the Green's function approach, and the method yields the energies and wave functions of the particle interacting with the impurities. Since the interaction
between the particles is effectively zero-ranged, the whole description of the problem is reduced to finding zeros of a simple function, given by a determinant of a
finite, known matrix.

The few-body system studied in this work is composed of a single atom and many impurities. We assume that each impurity is trapped tightly by its separate external
trapping potential. We consider the impurities to be localized at pre-determined positions and refer to them as static impurities. An example of such a
situation is displayed in Fig.~\ref{Fig:TIMechanism}, where a single atom moves in a harmonic trapping potential with two different impurities localized by separate traps
(purple and green potentials in the figure). The atom-impurity interaction is assumed to be local, i.e., the characteristic interaction range is much smaller than other
length scales such as the trap size and de Broglie wavelength.

The Hamiltonian of a trapped atom interacting with~$N$ static impurities is given by
\begin{equation}
\label{eqn:H_dim}
H = -\frac{\hbar^2}{2m}\Delta + V_{\mathrm{tr}}(\textbf{r}) + \sum_{i=1}^N V_{\mathrm{ai}}(\textbf{r}_i),
\end{equation}
where $V_{\mathrm{tr}}$ denotes the trapping potential, and $\textbf{r}_i = \textbf{r} - \textbf{d}_i$ is the position of the atom with respect to the $i$-th
impurity. The separation of length scales allows to approximate the true atom-impurity interaction potential by the contact pseudopotential
\begin{equation}
\label{eqn:FermiPseudopot}
  V_{\mathrm{ai}}(\mathbf{r}_i) = g_i\delta(\mathbf{r}_i)\frac{\partial}{\partial r_i}r_i.
\end{equation}
Here, the parameter $g_i = {2 \pi \hbar^2 a_i}/{m}$ is the coupling strength, which is expressed in terms of the effective atom-impurity scattering length $a_i$ and the
mass $m$ of the atom. Note that we allow the atom to interact with each impurity with its own potential, so the coupling strength $g_i$ can depend on the index $i$ of
the impurity.

The Hamiltonian from Eq.~\eqref{eqn:H_dim} leads to the following time-independent Sch\"odinger equation:
\begin{equation}
\label{eqn:SchrNIons}
  \bigg( - \frac{\hbar^2}{2m}\Delta + V(\mathbf{r}) + \sum_{i=1}^N g_i\delta(\textbf{r}_i)\frac{\partial}{\partial r_i}r_i\bigg)\Psi(\textbf{r}) = E
  \Psi({\textbf{r}}).
\end{equation}
In order to find the eigenstates of this equation, we start by expanding the (yet unknown) wave function $\Psi(\textbf{r})$ in the basis states $\phi_{\textbf{n}}(\mathbf{r})$, so that $\Psi(\textbf{r}) =
\sum_{\textbf{n}}c_{\textbf{n}} \phi_{\textbf{n}}(\textbf{r})$. We use the basis in which the noninteracting part of the Hamiltonian is diagonal.
To find the wave function $\Psi$, we insert its expansion in the chosen basis into Eq.~\eqref{eqn:SchrNIons} and obtain a set of equations for the coefficients $c_\mathbf{n}$ (see Appendix~\ref{app-sol} for details of the derivation). The result yields
\begin{equation}
\label{psi}
  \Psi(\textbf{r})= \sum_{i=1}^N \sum_{\textbf{n}}g_i k_i\frac{\phi_{\textbf{n}}^*( \textbf{d}_i) \phi_{\textbf{n}}(\textbf{r})}{E - E_{\textbf{n}}},
\end{equation}
where $k_i$ are given by
\begin{equation}
\label{eqn:kiDef}
  k_i =\bigg(\frac{\partial}{\partial r_i} r_i \Psi(\textbf{r})\bigg){\bigg|}_{\mathbf{r}= \mathbf{d}_i}.
\end{equation}
The solution $\Psi$ depends on the coefficients $k_i$, which in turn depend on $\Psi$. Therefore, solution has to be found in a self-consistent way.

To proceed with the construction of $\Psi$ and evaluation of $k_i$, we first recall the expression for the Green's function (see Appendix~\ref{app-green} fore more details):
\begin{equation}
\label{green}
  G(\mathbf{d}_i, \mathbf{r}) = \sum_{\textbf{n}}\frac{\phi_{\textbf{n}}^*( \textbf{d}_i) \phi_{\textbf{n}}(\textbf{r})}{E - E_{\textbf{n}}}.
\end{equation}
Note that $G$ depends on the energy E, but we dropped this dependence in the notation for brevity.
We then insert the solution for $\Psi$ from Eq.~\eqref{psi} into Eq.~\eqref{eqn:kiDef}, and rewrite it using Eq.~\eqref{green} to finally arrive at
\begin{equation}
\label{eqn:ki}
  k_i = \sum_{j=1}^N  g_j k_j\bigg(\frac{\partial}{\partial r_i} r_i G(\textbf{d}_j, \textbf{r})\bigg)\bigg|_{\textbf{r}=\textbf{d}_i}.
\end{equation}
This is a linear equation for the coefficients $k_i$, and it can be put into the matrix form
\begin{equation}
\label{fin}
\hat D_N \cdot \vec{k} = 0,
\end{equation}
where $\vec{k} = (k_1,\ldots, k_N)$ and the matrix
  \begin{equation}
    \label{eqn:Mk}
    \hat D_N(E)\! \!=\!\!
    \begin{pmatrix}
      \!g_1  G_\mathrm{r}(\textbf{d}_1, \textbf{d}_1)  \!-\!1&
      ... & g_N G(\textbf{d}_N, \textbf{d}_1)  \\
      \vdots & \ddots & \vdots
      \\
      g_1 G(\textbf{d}_1, \textbf{d}_N) &
      ...& g_N  G_\mathrm{r}(\textbf{d}_N, \textbf{d}_N)\!-\!1
    \end{pmatrix}\!,
  \end{equation}
where the regularized Green's function, which appears on the diagonal of $\hat D_N$, is $G_{\mathrm{r}}(\textbf{d}_i, \textbf{d}_i) =[\frac{\partial}{\partial r_i} r_i
  G_E(\textbf{d}_i, \textbf{r})]|_{\textbf{r}= \textbf{d}_i }$. Note, that the matrix $\hat D_N$ depends on the energy $E$ only through the Green's function.

Solutions of Eq.~\eqref{fin} exist provided that the determinant of $\hat D_N$ is equal to $0$.  For fixed positions $\textbf{d}_i$ and coupling strengths $g_i$, the
determinant is a function of a single variable $E$ only, and its roots are identified as the eigenenergies of the system, i.e., $\mathrm{det}\hat D_N(E_n)=0$ for the $n$-th
stationary state.  For each eigenenergy $E_n$, the corresponding wave function, expressed in terms of $k_i$, see Eq.~\eqref{psi}, is obtained by evaluating the kernel (the
null space) of the matrix~$\hat D_N(E_n)$.

\section{Two impurities in a harmonic trap}
\label{sec:Results}

With the general solution at hand, we now consider a single atom interacting with two impurities that are located at positions $\mathbf{d}_1$ and~$\mathbf{d}_2$ in a spherical harmonic trap
with frequency $\omega$.  We assume that all the scattering lengths are the same and equal to~$a$. To simplify the notation, we transform the problem into dimensionless units of oscillator length
and energy, $l_0 = \sqrt{{\hbar}/{m\omega}}$ and $E_0 = \hbar \omega$, respectively. The dimensionless coupling strength, naturally entering into the problem in
the place of the coupling $g_i=g$, is then equal to $\gamma=2 \pi a/l_0$.

\begin{figure}[H]
\includegraphics[width=0.45\textwidth]{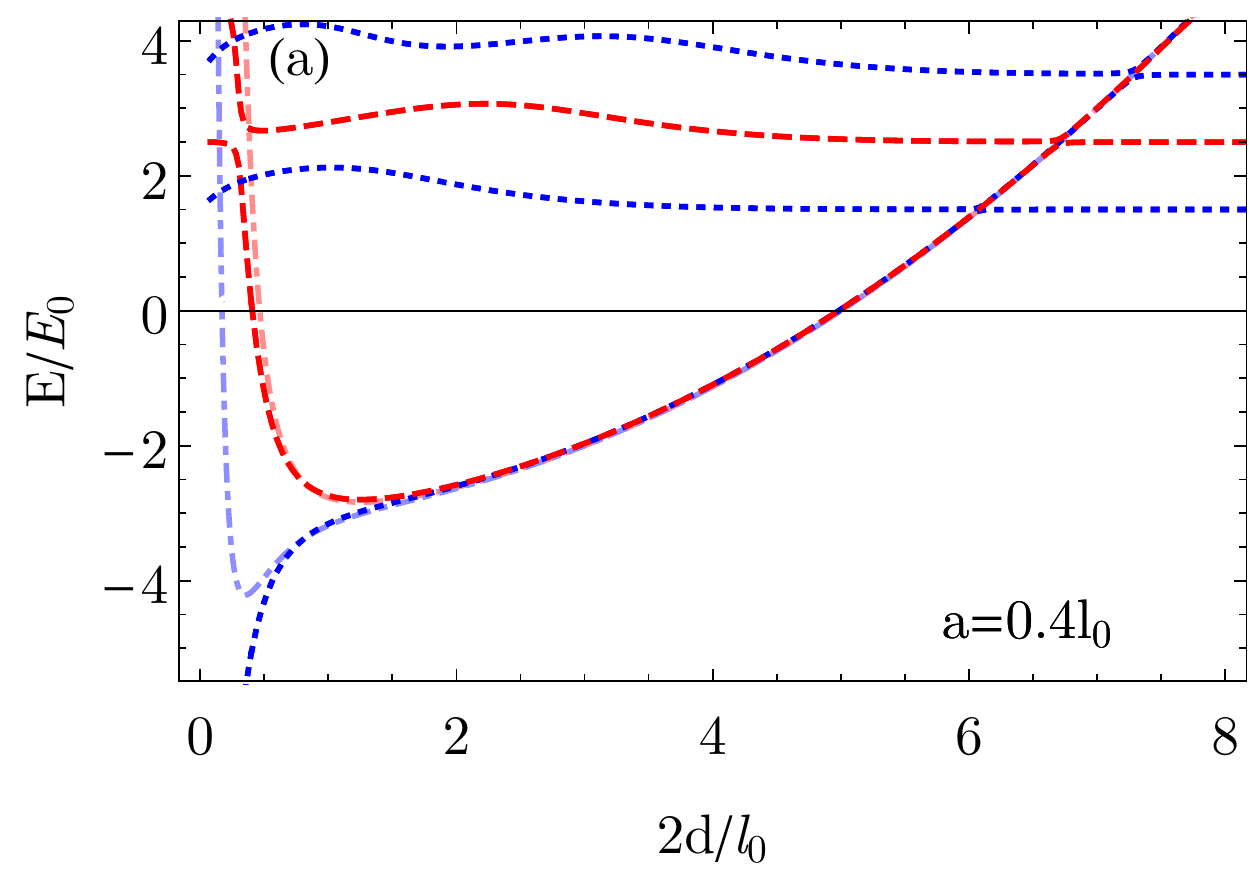}\\
\includegraphics[width=0.45\textwidth]{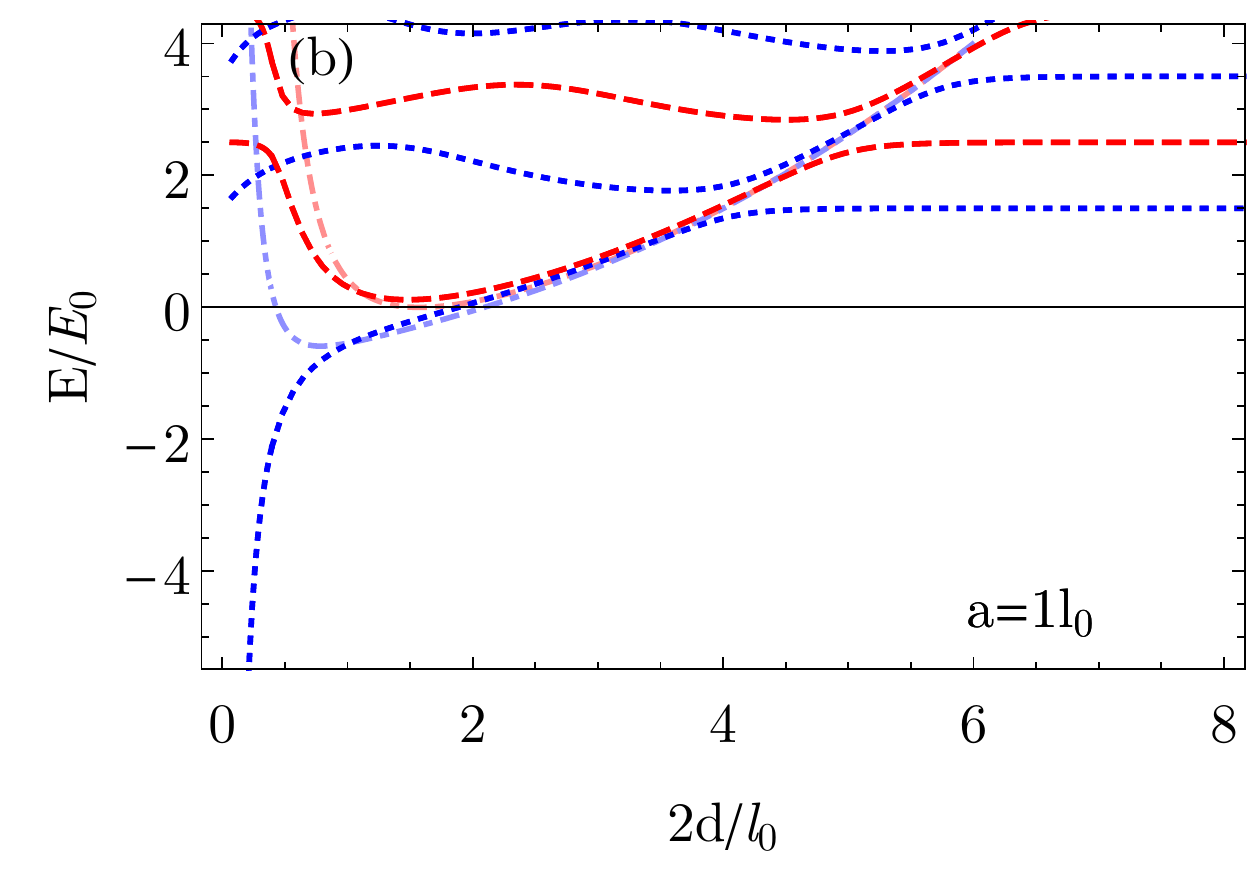}
\caption{(Color online) The dependence of the lowest energy levels on the distance $2d$ between the symmetrically placed impurities. The results are presented for
  $a=0.4\,l_0$ [upper panel (a)] and $a=l_0$ [lower panel (b)]. The color code of the levels is the same as in Fig.~\ref{fig:energyLevels}. Additionally, the dot-dashed
  lighter blue and lighter red lines represent the energies of the even and odd bound states, respectively, obtained within variational
  approach.} \label{Fig:groundStatesComp}
\end{figure}

The stationary states, their energies and wave functions are calculated from Eq.~\eqref{fin}. Here, $\vec{k} = (k_1,k_2)$, and the matrix $\tilde D(E) \equiv
\hat D_2(E)/\gamma$ stems from Eq.~\eqref{eqn:Mk}:
\begin{equation}
\label{tD}
  \tilde D(E) = \begin{pmatrix}
    G_{\textrm{r}}(\textbf{d}_1, \textbf{d}_1)  -\gamma^{-1} & G(\textbf{d}_2, \textbf{d}_1)  \\
    G(\textbf{d}_1, \textbf{d}_2) & G_{\textrm{r}}(\textbf{d}_2, \textbf{d}_2)-\gamma^{-1}
  \end{pmatrix}.
\end{equation}

Since in the matrix $\tilde D(E)$ the rows and columns are linearly dependent, only the ratio of $k_i$ can be evaluated, i.e., $k_1/k_2 = - [\tilde D(E)]_{12} / [\tilde
D(E)]_{11}$. The absolute values of $k_i$ can then be determined from the normalization condition for $\Psi$ in Eq.~\eqref{psi}.

\subsubsection*{Symmetric case}
Below we focus on the case of two impurities placed symmetrically with respect to the origin, $\textbf{d}_1 =
-\mathbf{d}_2 = \textbf{d}$. We assume that the impurities are located on the $z$-axis, and we take $\mathbf{d}=(0,0,d)$. Note, that the distance between the impurities is equal to $2d$.

\begin{figure}[H]
  \begin{subfigure}[b]{\textwidth}
    \includegraphics[width=0.45\textwidth]{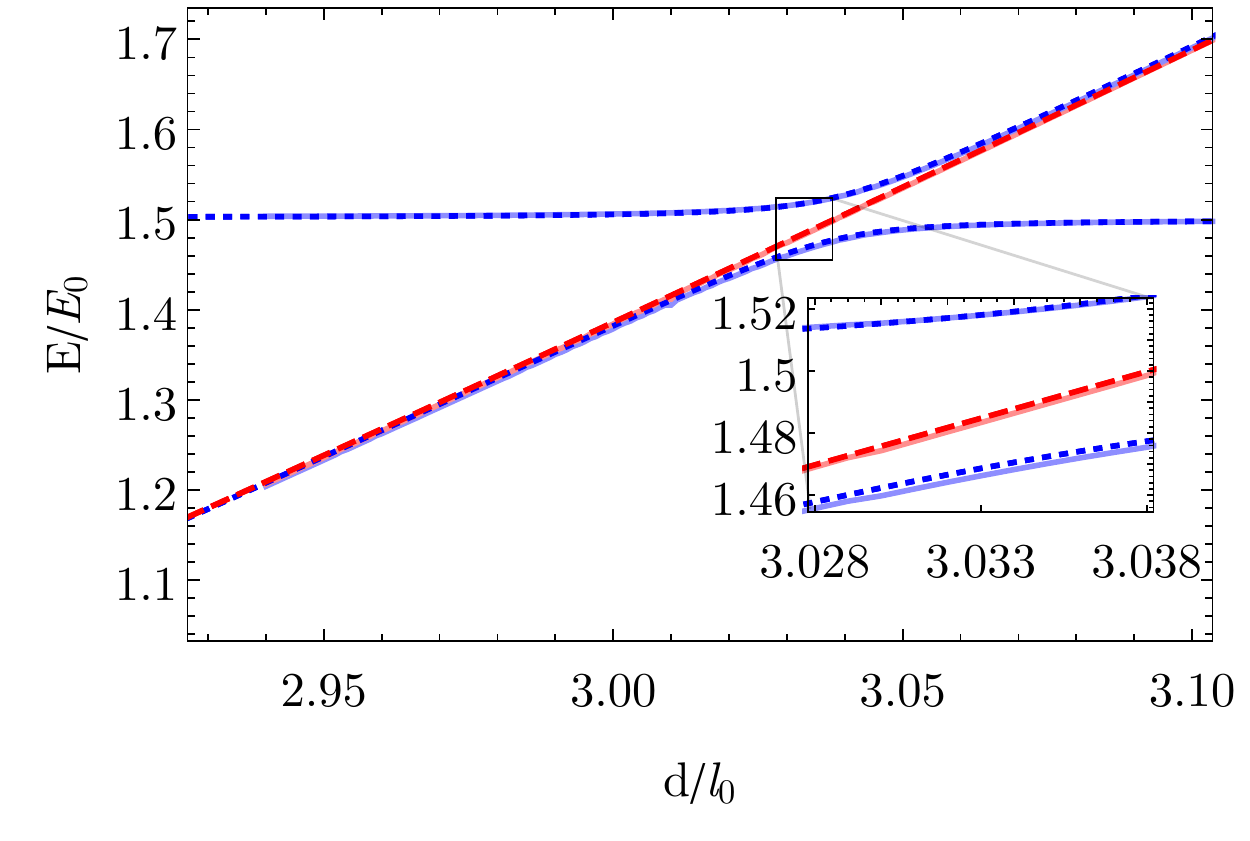}
  \end{subfigure}
  \caption{(Color online) Comparison of the results for $a=0.4\,l_0$ in the vicinity of the lowest avoided crossing obtained within the full method (dotted blue and
    dashed red), and within the variational approach using three states (lighter blue and lighter red). The inset zooms the vicinity of the avoided crossing.  The color
    code is the same as in Fig.~\ref{Fig:groundStatesComp}.} \label{fig:avoided}
\end{figure}

To find the energies, we search for the roots of $\det \tilde D(E)$ given by Eq.~\eqref{tD}.  First, we calculate the energies of the system for different
values of the distance $2d$ between the impurities and different scattering length $a$ characterizing the atom-impurity interaction.

Fig.~\ref{fig:energyLevels} presents the dependence of the energy levels of the system on the impurities' positions for six different values of $a/l_0 = \pm0.4$, $\pm
1.0$, and $\pm10$.  In general, the eigenstates can be classified according to the symmetry $z\to-z$ of the Hamiltonian into even and odd states, denoted in the figure
with blue dotted and red dashed lines, respectively. As can be observed from the figure, for very large separations between the impurity atoms, the energy spectrum
approaches the spectrum of the unperturbed harmonic oscillator, $E_n^\mathrm{ho} = \hbar\omega(n + 3/2)$ with $n=0,2,4,\ldots$ for even and $n=1,3,5,\ldots$ for odd
states.  For separations comparable to the oscillator length, the observed energies deviate from the harmonic oscillator case due to the presence of the impurities.  For
separations between the impurities much smaller than the other length scales of the model ($a$ and $l_0$), when the distance~$d$ is of the order of the interaction range
of the true potential, the description of the interaction in terms of the contact pseudopotential is no longer valid. Interestingly, the odd states do not feel the
contact potential for $d=0$, recovering the unperturbed harmonic oscillator limit in this case, but the even states for $d=0$ do not approach the results obtained by
Busch~\cite{busch1998} for a single impurity. In the limit $d \to 0$, our model in terms of two separate regularized delta potentials is no longer valid.

Let us first discuss the results for negative~$a$ presented in the bottom row in Fig.~\ref{fig:energyLevels}.  In the case of $a=-0.4l_0$ (see
Fig.~\ref{fig:energyLevels}d), we observe relatively small perturbation compared to the harmonic oscillator case. The energy shift becomes larger with increasing
magnitude of the scattering length $a$ (see Figs.~\ref{fig:energyLevels}e--f). However, for small~$d$, when the harmonic potential is negligible, the energy of the atom
is negative, indicating the presence of a bound state.

To identify the lowest energies of the atom with bound states for small $d$, we calculate the bound state energies of the atom in free space, with neglected trapping
potential.  To this end, we refer to Eq.~\eqref{tD}, when now $G$ denotes the Green's function of the atom in free space. The results, i.e., the roots of $\det
\tilde{D}(E)$ with $E<0$, are depicted in Fig.~\ref{fig:energyLevels} with gray solid lines. In free space, for $a<0$, only even bound states (with $E<0$) exist if $2d <
|a|$. At $2d=|a|$ the energy of the state crosses the zero threshold and enters into the continuum.

Let us now turn to the positive values of the scattering length $a$ (see Fig.~\ref{fig:energyLevels}a). At large separation between the impurities, the lowest state is a
doubly degenerate superposition of dimer bound states.  The energy of the dimer in free space approaches $- \hbar^2/2m a^2$, but it is lifted quadratically in our case
due to the external harmonic trap.  At small separations, the trapping potential is negligible and the splitting between the bound states of different symmetry becomes
significant.  The state lower in energy is always even and the higher is odd. For larger scattering lengths, the odd state can even be pushed into the continuum (compare
Figs.~\ref{fig:energyLevels}b and \ref{fig:energyLevels}c). We note that the splitting between the bound states decays exponentially with the distance, as is the case for
the H$_2^+$ molecule~\cite{Guo1993}. The system considered here acts as a precursor for molecular physics simulation by reproducing the core features of the simplest possible molecule.

In general, avoided crossings appear due to the trap-induced shape resonance mechanism~\cite{Stock2003}.
Each of the potentials, describing impurity-atom interaction, can support a bound state. Its energy can be lifted above the zero energy threshold by the external
potential.  If the total energy 
is brought into degeneracy with this bound state,
a the trap-induced resonance occurs. This is similar to the simpler case of two harmonically trapped atoms studied
in~\cite{Krych2009}, but more complex due to the reflection symmetry present in our problem. Trap-induced resonances can be used to generate entangled states and perform
gate operations~\cite{Stock2003,Krych2009}.

In Fig.~\ref{fig:energyLevels}a, we observe narrow avoided crossings since the scattering length is small, and thus the coupling between the levels is weak.  The splitting in
the avoided crossings increases with growing $a$, as can be observed from Figs.~\ref{fig:energyLevels}b and~\ref{fig:energyLevels}c.  The reflection symmetry of the system with respect to
$z\to-z$ implies the presence of a state of different symmetry between each two states of the same symmetry experiencing an avoided crossing. Specifically, at the
positions of the avoided crossings in Fig.~\ref{fig:energyLevels}a--c, between each two energy levels of the same symmetry (the same color) there is a state of different symmetry (of
other color). This effect originates from the presence of two bound states, which are almost degenerate, but have different symmetries and do
not couple with each other.

\begin{figure}[H]
  \begin{subfigure}[b]{\textwidth}
    \includegraphics[width=0.46\textwidth]{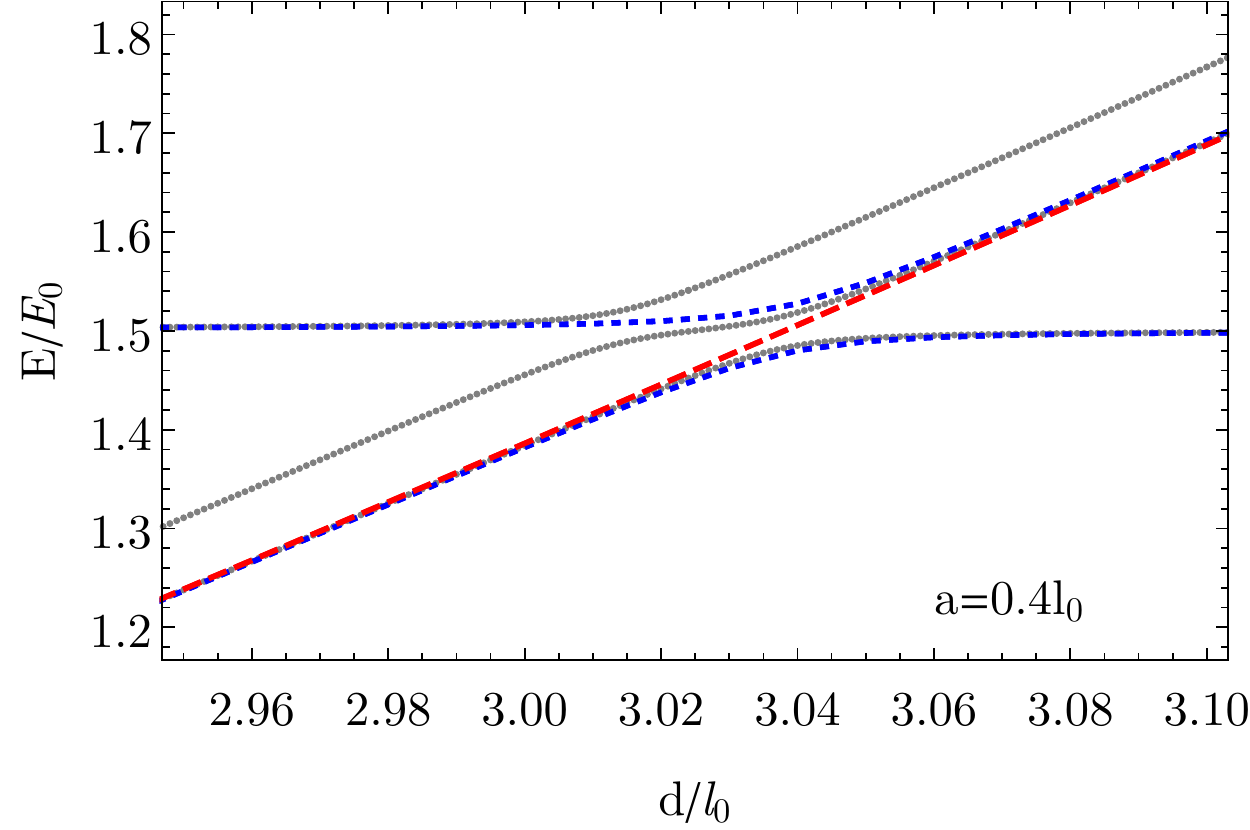}
  \end{subfigure}
  \caption{(Color online) Comparison of energy levels in the vicinity of avoided crossing. Red dashed line and blue dotted line denote odd and even states of the system with
    impurities placed symmetrically along the $z$-axis in $z=\pm d$. Grey dotted line denotes the energy levels of the system, where the impurities are placed in $z=-d$
    and $z=d+\Delta d$. Here $\Delta d = 0.025 l_0$.}\label{Fig:ACnonsymmmetric}
\end{figure}

To understand the properties of the avoided crossing between the extended states and the bound states in trap, we turn to a simpler description. We adopt the variational approach in which we make the following ansatz:
\begin{equation}
\label{ansatz}
  \Psi(\mathrm{r}) = v_1 \psi_1(\mathbf{r})+v_2 \psi_2(\mathbf{r}),
\end{equation}
where the normalized wave function $\psi_i(\mathbf{r}) = \psi(\mathbf{r} - \mathbf{d}_i)$, and the free-space wave function of the bound state is $\psi(\mathbf{r}) =
\exp(-r/a)/(\sqrt{2\pi a} r)$. Therefore, the wave function is a linear combination of states that describe an atom localized around each impurity. The minimum of the energy is achieved for $v_i$ that satisfy
\begin{equation}
  \begin{pmatrix}
    H_{11} - E & H_{12} -ES_{12}\\
    H_{21} - ES_{21} & H_{22} - E
  \end{pmatrix}
  \begin{pmatrix}
    v_1\\v_2
  \end{pmatrix} = 0,
\end{equation}
where the matrix elements of the Hamiltonian are denoted by $H_{ij} =\braket{\psi_i|H|\psi_j}$, and the overlap between the states is $S_{ij} =
\braket{\psi_i|\psi_j}$. We solve the resulting equations numerically.

The results of the variational approach are presented in Fig.~\ref{Fig:groundStatesComp}. The bound states of different symmetries (for odd states we have $v_1 = -v_2$ whereas
for even $v_1=v_2$) are displayed for positive scattering lengths $a=0.4\, l_0$ and $a=\l_0$ in Figs.~\ref{Fig:groundStatesComp}a and~\ref{Fig:groundStatesComp}b,
respectively.  Even though the approximate wave function works well reproducing the overall trend, the approach is missing the quantitative description of the avoided
crossings. Furthermore, the approximation breaks down when the distance between the impurities is comparable to the scattering length, and the overlap $S_{12}$ deviates
significantly from zero. In all the other cases, i.e., for larger impurity separations and away from the avoided crossing, the variational calculation is accurate.

To improve the approximate description of the wave function in variational approach, we include into Eq.~\eqref{ansatz} a third state, which corresponds to an extended
state (occupying the whole volume of the trap) of the unperturbed harmonic oscillator.  For illustration, we will only consider the lowest trap-induced shape resonance, which occurs for $a/l_0=0.4$ at $2d/l_0
\approx 6$.  To this end, we add a third state $\psi_3(\mathbf{r}) = \phi_0(\mathbf{r})$, where $\phi_0(\mathbf{r}) \propto \exp(-r^2/2l_0^2)$ is the normalized ground
state wave function of the harmonic oscillator, with its corresponding amplitude $v_3$ on the right-hand side in Eq.~\eqref{ansatz}. The minimization of the mean energy with such an ansatz
yields the energy as a function of the distance $2d$ between the impurities.

In Fig.~\ref{fig:avoided} we show the zoom in of the avoided crossing for $a/l_0=0.4$. The full, original results are depicted with dotted blue and dashed red
curves, whereas lighter blue and lighter red colors are dedicated for the variational approach. Clearly, since the curves obtained within different methods collapse onto each other, the
simple three-state model gives the quantitative description of the trap-induced resonance. Notice the presence of the state of different symmetry which passes through the avoided crossing
(red dashed straight line) without being affected by the other states.

\subsubsection*{Asymmetric case}

The states and the energy levels of the atom divide into separate classes, belonging to different irreducible representations of
the symmetry group~\cite{landau1958course}, characterized by different symmetry properties. Since $\mathbf{d}_1 = - \mathbf{d}_2$ the Hamiltonian
is invariant with respect to the reflection in the plane passing in between the impurities and perpendicular to the line joining the particles.  To see how the coupling
between the states affects the energy levels, we break the symmetry by perturbing one impurity's position. Now, the position $\mathbf{d}_2 = -d \mathbf{e}_z$ is unaffected, whereas
$\mathbf{d}_1 = (d + \Delta z) \mathbf{e}_z $, where we denote by $\mathbf{e}_z$ the unit vector pointing along the $z$-axis.


\begin{figure*}
  \begin{subfigure}[b]{0.32\textwidth}
    \includegraphics[width=\textwidth]{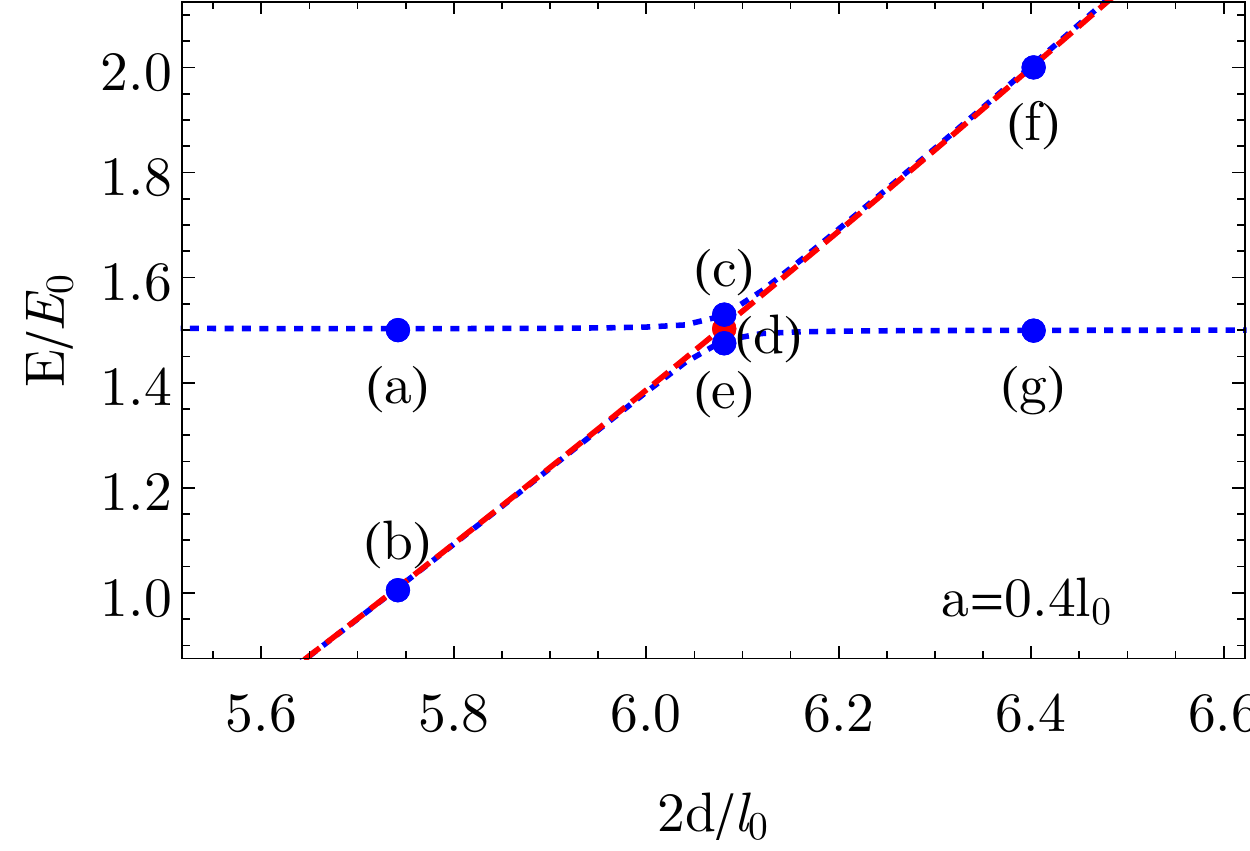}
  \end{subfigure}
  \begin{subfigure}[b]{0.32\textwidth}
    \includegraphics[width=\textwidth]{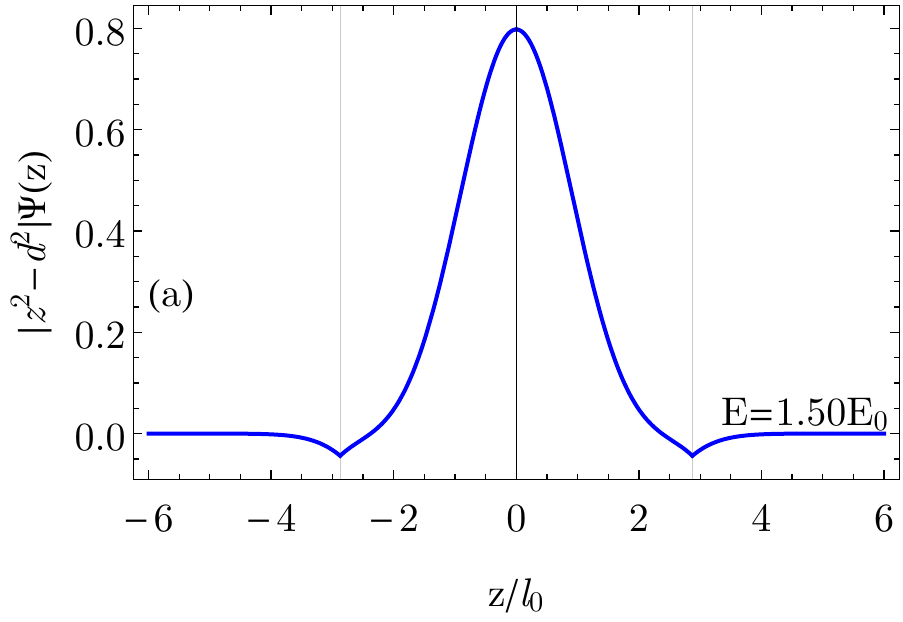}
  \end{subfigure}
  \begin{subfigure}[b]{0.32\textwidth}
    \includegraphics[width=\textwidth]{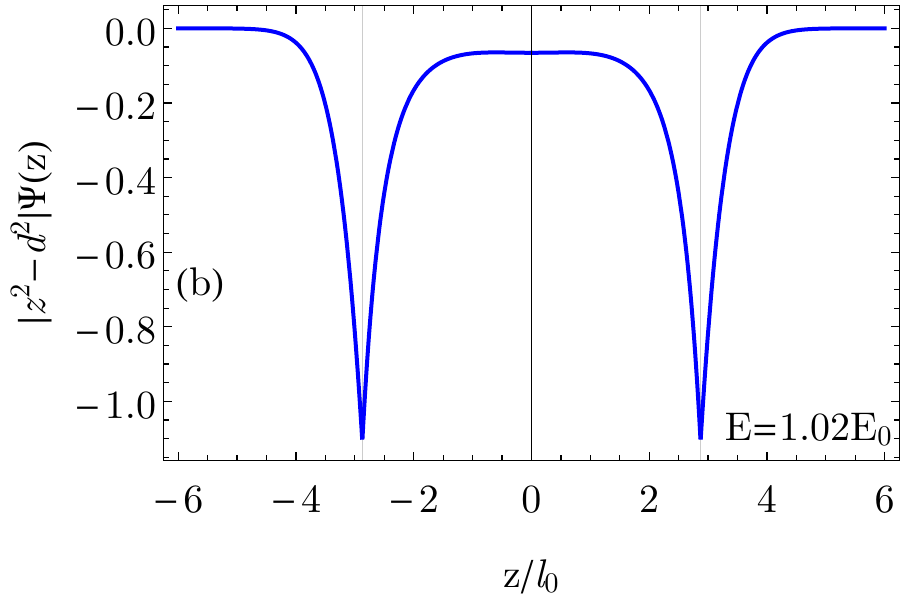}
  \end{subfigure}\\
  \begin{subfigure}[b]{0.32\textwidth}
    \includegraphics[width=\textwidth]{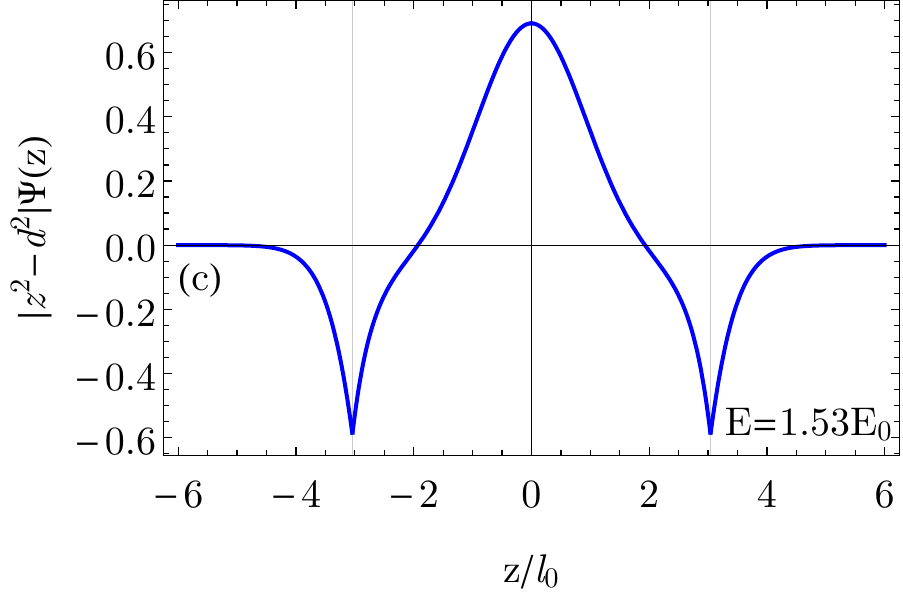}
  \end{subfigure}
  \begin{subfigure}[b]{0.32\textwidth}
    \includegraphics[width=\textwidth]{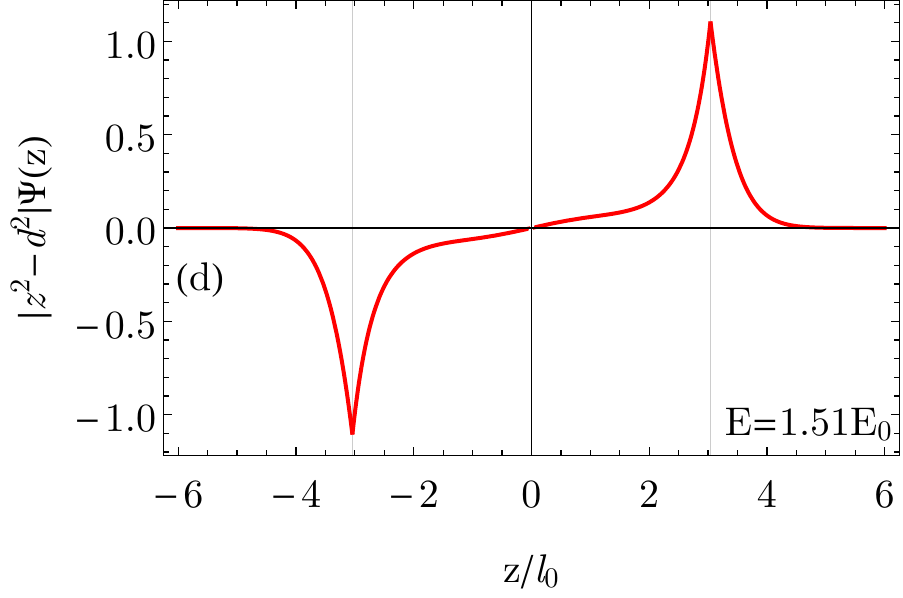}
  \end{subfigure}
  \begin{subfigure}[b]{0.32\textwidth}
    \includegraphics[width=\textwidth]{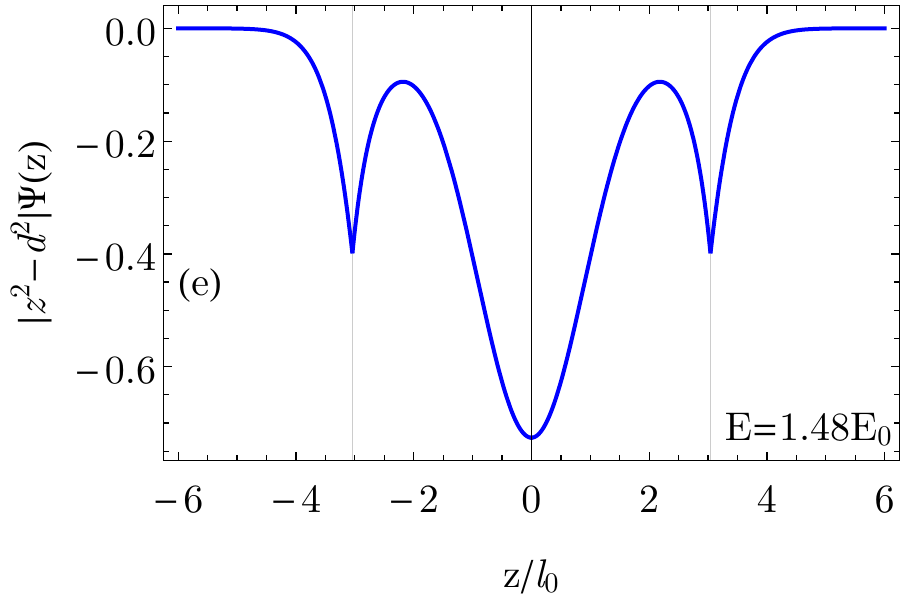}
  \end{subfigure}\\
  \begin{subfigure}[b]{0.32\textwidth}
    \includegraphics[width=\textwidth]{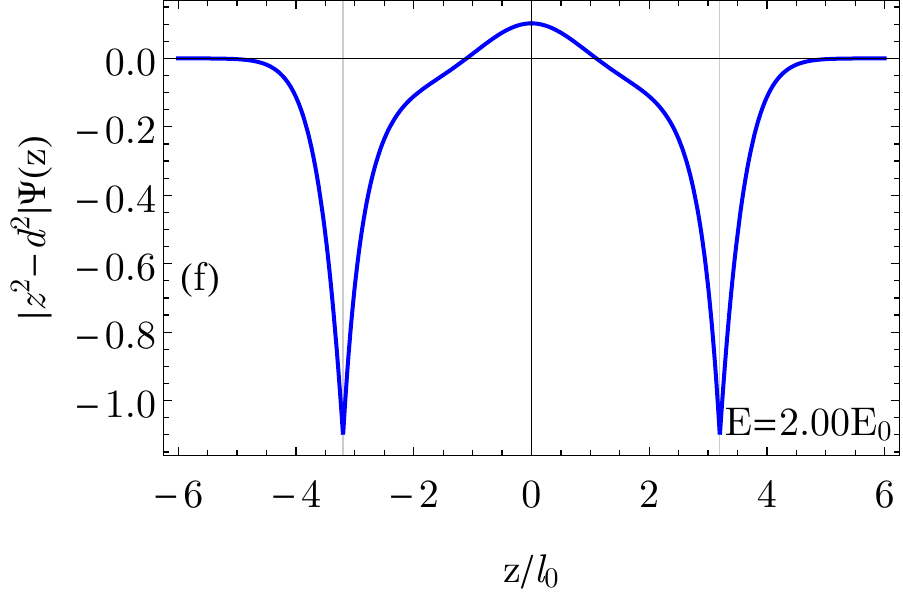}
  \end{subfigure}
  \begin{subfigure}[b]{0.32\textwidth}
    \includegraphics[width=\textwidth]{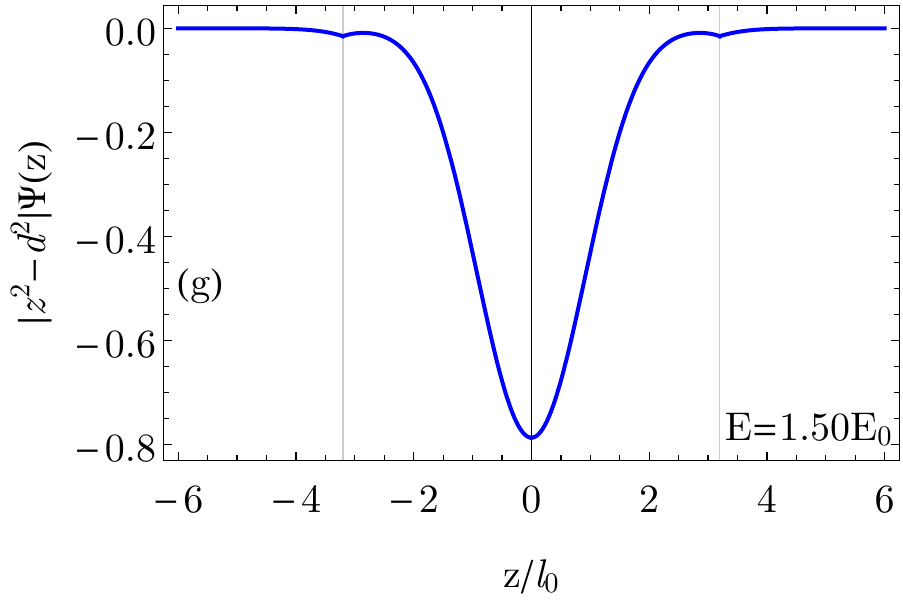}
  \end{subfigure}
  \begin{subfigure}[b]{0.32\textwidth}
    \includegraphics[width=\textwidth]{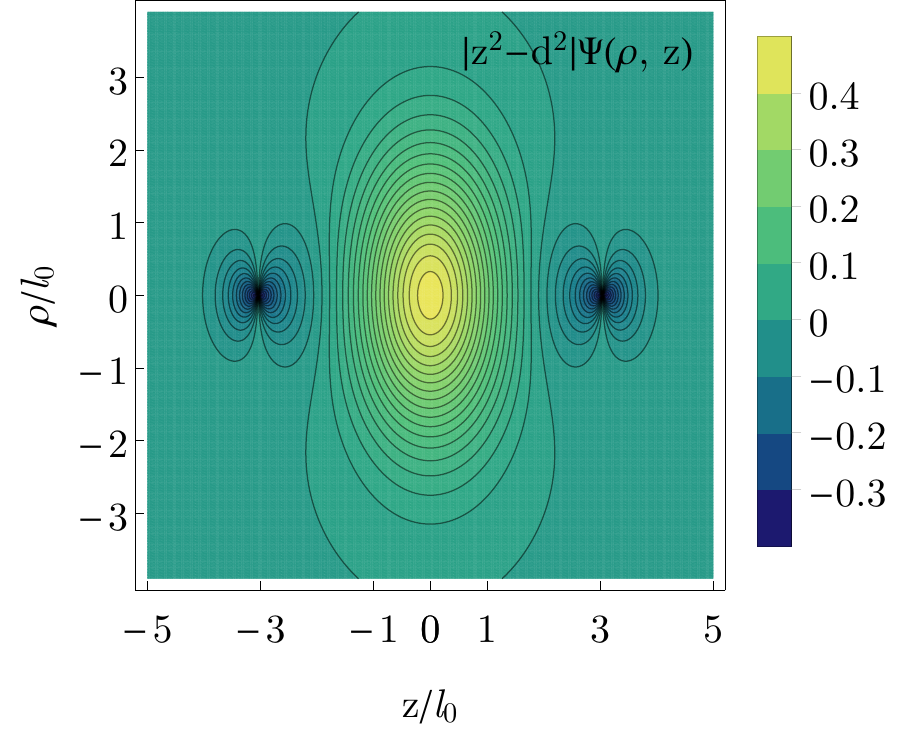}
  \end{subfigure}
  \caption{(Color online) Cuts along the $z$-axis of the (renormalized) wave functions of the atom for different $d$ and $E$ with the scattering length a = 0.4$l_0$ close to avoided
    crossings. Gray vertical lines denote the positions of the ions. The last picture shows the contour plot of the wave function presented in (c).}\label{fig:avoidedCrossing}
\end{figure*}


The energy levels of the atom in such a configuration with $\Delta z = 0.025 l_0$ are presented in Fig.~\ref{Fig:ACnonsymmmetric}. The dotted blue (even states) and dashed red (odd states) lines are
the full solutions of the initial, unperturbed system, whereas the small-dotted gray line represents energy levels of the perturbed Hamiltonian. All the states are
repelling, lifting the degeneracy, which results in two very close avoided crossings between these states and one of the extended state in the harmonic trap. This
twin-resonance, facilitated by the controlled symmetry breaking of the system and by the presence of the trap, signals the breakdown of the usual Landau-Zener
theory~\cite{carroll1986generalisation,carroll1985further,shytov2004landau}.

\subsubsection*{Wave functions in the symmetric case of two impurities}

With our method we also determine the wave function of the atom. Provided the coefficients $k_i$ are known, the wave function is evaluated from Eq.~\eqref{psi}, and it
takes the form
\begin{equation}
  \Psi(\mathbf{r}) = \sum_{i=1}^N g_i k_i G(\mathbf{d}_i,\mathbf{r}).
\end{equation}
In this sum, the energy $E$ as well as the eigenstates are determined from Eq.~\eqref{fin}. So far, we considered the energy levels of the atom, and therefore we already determined the matrix $\hat D_2(E)$, which in our
case of two symmetrically placed impurities takes a dimensionless form of $\tilde D(E)$, see Eq.~\eqref{tD}. The solution is then particularly simple since the symmetry
property imposes $k_1 = k_2$ for even states, and $k_1=-k_2$ for odd states.

In Fig.~\ref{fig:avoidedCrossing}, we present the cuts along the $z$-axis of the wave functions of seven eigenstates in the vicinity of the lowest avoided crossing for
$a=0.4\,l_0$. For the clarity of presentation, we plot the wave functions multiplied by a factor $|z^2-d^2|$ to remove the divergence, which appears for $z=\pm d$ and
$x=y=0$. Each divergence originates in the Green's function, which has a pole when its two arguments approach each other, i.e., $G(\mathbf{d}_i,\mathbf{r}) \propto
1/|\mathbf{r} - \mathbf{d}_i|$ for $\mathbf{r} \approx \mathbf{d}_i$. This divergence is responsible for the limiting behavior of the wave function at vanishing
atom-impurity distance. According to the contact condition, $\Psi(\mathbf{r})$ is proportional to $1- a/|\mathbf{r} - \mathbf{d}_i|$ in this case.

The first plot (upper left corner of the panel) in Fig.~\ref{fig:avoidedCrossing} magnifies the relevant avoided crossing that we will investigate here in more
details. The seven points, marked with letters a--g, indicate the parameter values for which the eigenstates are studied on further plots.  The point~(a) indicates the
trap extended, symmetric state of the atom with small admixture of the states localized on the impurities. In Fig.~\ref{fig:avoidedCrossing}b, corresponding to point~(b),
the symmetric bound state of the atom is shown. According to the variational model, defined by Eq.~\eqref{ansatz}, these two states are mixed in the vicinity of the
avoided crossing and corresponding amplitudes, $v_1$ and $v_2$, are of the same order. This is shown in Figs.~\ref{fig:avoidedCrossing}c and~\ref{fig:avoidedCrossing}e,
in which the two states corresponding to~(c) and~(e) are indeed mixed, with the trap-extended and localized components of the wave function clearly visible. In between
these two states (in energy), one finds the localized wave function of the bound state with odd symmetry, see Fig.~\ref{fig:avoidedCrossing}d corresponding to the
point~(d). When the separation between the impurities is increased further, the extended and the bound states are again weakly coupled, as can be seen from
Figs.~\ref{fig:avoidedCrossing}f, a bound state corresponding to~(f), and~\ref{fig:avoidedCrossing}g, a trap extended state corresponding to~(g). For completeness, we
also present here a two-dimensional cut of the wave function along $x$- and $z$-axes. This wave function, corresponding to the case~(c), is also presented in
Fig.~\ref{fig:avoidedCrossing} (the bottom right corner of the panel).

\section{Summary}
\label{sec:Conc}
In this work, we presented a general method of solving the problem of a single atom interacting with $N$ stationary impurities.  The approach is based on the Green's
function formalism, and assumes the contact potential approximation. The method can be applied for arbitrary arrangement of the impurities, even when the interaction
strength is different for each one.

We applied the method to the case of two impurities placed in a spherical harmonic trap. We determined energies and wave functions of stationary states of the
atom. The spectrum exhibits multiple avoided crossings between the bound states and the extended trap states. A simple three-states model correctly reproduces the bound
states in the trap as well as the trap-induced resonances.

Our results can be further generalized to include energy-dependent scattering lengths, which would allow for more accurate treatment of long-range potentials, for
instance the atom-ion polarization potential~\cite{melezhik2016confinement}. The method, by providing single particle orbitals, can serve as a starting point for more
involved calculations, such as dynamics of the atom in complex quantum networks of impurities, or many-body system of weakly interacting bosons interacting with multiple
trapped ions~\cite{schurer2016impact}. It is possible to include motion of the impurities within the method, possibly capturing effects such as atom-phonon coupling.

This work presents a study of a simplified case in which the atom-impurity interaction is described using a zero-range potential. This is sufficient as long as the
characteristic length scale of the interaction is much smaller than other length scales such as the interparticle distance. Within this treatment the system has some
characteristic features of a diatomic molecule such as the presence of even and odd states. However, the truly interesting case would be the one when the atom interacts
strongly with many impurities at the same time, where the zero-range model does not apply. Experimental realization of such a system would require bringing the impurities
within the characteristic atom-impurity interaction distance, e.g., hundreds of nanometers in the ion-atom case. This cannot currently be achieved with stationary
impurities. Rigorous theoretical description of such a system would require including the motion of the impurities as well as using realistic interaction potentials,
resulting in a numerically challenging problem. The current results can then serve as a limiting case.

\section*{Acknowledgements}
We thank Antonio Negretti and Rene Gerritsma for valuable discussions. This work was supported by the Polish National Science Center projects 2014/14/M/ST2/00015 and DEC-2013/09/N/ST2/02188 and the Alexander von Humboldt Foundation.

\appendix
\section{Solution of the Schr\"{o}dinger equation}
\label{app-sol}
In order to solve the Schr\"{o}dinger equation for the atom (see Eq.~\eqref{eqn:SchrNIons}), we first expand the unknown wave function $\Psi(\textbf{r}) =
\sum_{\textbf{n}}c_{\textbf{n}} \phi_{\textbf{n}}(\textbf{r})$ in the basis $\phi_\mathbf{n}$ of the stationary states of the atom but without the impurities.  Inserting
the expansion of $\Psi(\textbf{r})$ into Eq.~\eqref{eqn:SchrNIons}, we obtain
\begin{flalign}
  \begin{split}
    \sum_{\textbf{n}}c_{\textbf{n}} E_{\textbf{n}}\phi_{\textbf{n}}(\textbf{r}) + \sum_{i=1}^Ng_i\delta(\textbf{r}_i) \frac{\partial}{\partial r_i} r_i \left(
    \sum_{\textbf{n}}c_{\textbf{n}}\phi_{\textbf{n}}(\textbf{r})\right) =\\= E \sum_{\textbf{n}}c_{\textbf{n}} \phi_{\textbf{n}}(\textbf{r}),\label{eqn:Schr2}
  \end{split}
\end{flalign}
where $E_{\textbf{n}}$ denotes the energy corresponding to the state~$\phi_\mathbf{n}$.  The next step is to project both sides of Eq.~(\ref{eqn:Schr2}) onto a single
state of the basis $\phi_{\textbf{m}}^*$, in order to determine the expansion coefficients $c_{\textbf{m}}$:

\begin{equation}
  \sum_{i=1}^N g_i \phi_{\textbf{m}}^*( \textbf{d}_i) \frac{\partial}{\partial r_i} r_i
  \bigg(\sum_{\textbf{n}}c_{\textbf{n}}\phi_{\textbf{n}}(\textbf{r})\bigg)_{\textbf{r}\to\textbf{d}_i} \!\!\!\!\!= (E \!-\! E_{\textbf{m}}) c_{\textbf{m}}
\end{equation}
Now, we replace back the expansion $\sum_{\textbf{n}}c_{\textbf{n}} \phi_{\textbf{n}}(\textbf{r})$ with~$\Psi(\textbf{r})$:
\begin{flalign}
\begin{split}
c_{\textbf{m}} (E - E_{\textbf{m}})= \sum_{i=1}^Ng_i\phi_{\textbf{m}}^*( \textbf{d}_i) \bigg(\frac{\partial}{\partial r_i} r_i \Psi(\textbf{r})\bigg)_{\textbf{r}\rightarrow \textbf{d}_i} . \label{eqn:cmcp1}
\end{split}
\end{flalign}
Dividing both sides of Eq.~(\ref{eqn:cmcp1}) by $(E - E_\textbf{m})$, we finally obtain the equation for the expansion coefficients~$c_\textbf{m}$:
\begin{flalign}
\begin{split}
c_{\textbf{m}} = \sum_{i=1}^N g_ik_i \frac{\phi_{\textbf{m}}^*( \textbf{d}_i)}{(E - E_{\textbf{m}})}   ,  \label{eqn:cpcm2}
\end{split}
\end{flalign}
where
\begin{equation}
k_i =\bigg(\frac{\partial}{\partial r_i} r_i \Psi(\textbf{r})\bigg)_{\textbf{r}\rightarrow \textbf{d}_i}. \label{eqn:kiApp}
\end{equation}
Substituting Eq.~(\ref{eqn:cpcm2}) into the expansion of $\Psi(\textbf{r})$ yields the wave function in the following form:
\begin{equation}
\Psi(\textbf{r})=  \sum_{i=1}^N  \sum_{\textbf{n}}g_i k_i\frac{\phi_{\textbf{n}}^*( \textbf{d}_i) \phi_{\textbf{n}}(\textbf{r})}{(E - E_{\textbf{n}})} = \sum_i k_i  G(\textbf{d}_i, \textbf{r}), \label{eqn:psi}
\end{equation}
where $G(\textbf{d}_i, \textbf{r}) = \sum_{\textbf{n}}{\phi_{\textbf{n}}^*( \textbf{d}_i) \phi_{\textbf{n}}(\textbf{r})}/{(E - E_{\textbf{n}})}$ is the Green's function
(see Appendix~\ref{app-green} for details), and, therefore, we arrive at the consistency condition given by
\begin{equation}
\label{eqn:kikjApp}
k_i= \sum_{j=1}^N g_j k_j\bigg(\frac{\partial}{\partial r_i} r_i G(\textbf{d}_j, \textbf{r})\bigg)_{\textbf{r}\rightarrow \textbf{d}_i}.
\end{equation}
Here, we notice that in the case of $i\neq j$ the regularization operator is redundant, i.e., $\big(\frac{\partial}{\partial r_i} r_i G(\textbf{d}_j,
\textbf{r})\big)_{\textbf{r}\rightarrow \textbf{d}_i} = G(\textbf{d}_i, \textbf{d}_j)$. Therefore, the condition in Eq.~\eqref{eqn:kikjApp} can be rewritten in the form
of $\hat D_N(E)\cdot \vec{k}=0$, with $\hat D_N(E)$ given by Eq.~\eqref{eqn:Mk}.

\section{Green's function for spherically symmetric harmonic potential}
\label{app-green}

We discuss here the properties of the Green's function for an isotropic 3D harmonic oscillator. The analytical formulas for $n$-dimensions were found
in~\cite{Bakhrakh1972}. In the case of an anisotropic harmonic trap, the Green function can be expressed in terms of an integral that has to be calculated
numerically~\cite{Idziaszek2005}.  The Green's function of a system described by the Hamiltonian $H_0$ is defined by
\begin{equation}
  (H_{\mathrm{0}}-E)G (\textbf{r}, \textbf{r}')  = -\delta(\textbf{r}-\textbf{r}').
\end{equation}
This equation can be solved by expanding $G$ in the basis of $H_{\mathrm{0}}$, i.e., 3D harmonic oscillator wave functions in our case, and the final expression is
\begin{equation}
\label{ge}
  G(\mathbf{r}', \mathbf{r}) = \sum_{\textbf{n}}\frac{\phi_{\textbf{n}}^*( \textbf{r}') \phi_{\textbf{n}}(\textbf{r})}{E - E_{\textbf{n}}},
\end{equation}
where $\phi_\mathbf{n}$ is the eigenfunction of $H_\mathrm{0}$ with eigenvalue $E_\mathbf{n}$. This expression is exactly the one in Eq.~\eqref{green}.

The Green's function of the isotropic harmonic oscillator was calculated analytically
in~\cite{macek1998}, and is given in terms of the confluent hypergeometric functions~$U$ and~$M$:
\begin{widetext}
  \begin{flalign}
    \begin{split}
      G(\textbf{r}, \textbf{r}') &= \exp{\bigg(\!\!-\frac{\xi + \eta}{2}\bigg)}\bigg\{\Lambda(1, E)\bigg(1 + \frac{2 \xi \eta}{\xi - \eta}
      \bigg(\frac{\partial}{\partial \eta} - \frac{\partial}{\partial \xi} \bigg)\bigg) U_E^{(1)}(\xi)M_E^{(1)}(\eta) +\\ & +\mathrm{sign}(\textbf{r}\cdot
      \textbf{r}')\Lambda(1, E + 1)\frac{2\sqrt{\xi\eta}}{\xi-\eta}\bigg(\eta\frac{\partial}{\partial \eta} - \xi\frac{\partial}{\partial \xi}
      \bigg)U_{E+1}^{(1)}(\xi)M_{E+1}^{(1)}(\eta) \bigg\},
    \end{split}
    \label{eqn:Greens1}
  \end{flalign}
\end{widetext}
where the function $\Lambda$ is expressed in terms of the Euler gamma function,
\begin{equation}
  \Lambda(1, E) = -\frac{1}{2 }\bigg(\frac{1}{\pi}\bigg)^{3/2}\Gamma\bigg(\frac{3}{4} - \frac{E}{2} \bigg),
\end{equation}
while the dimensionless parameters $\xi$ and $\eta$ depend on the positions $\mathbf{r}$ and $\mathbf{r}'$:
\begin{eqnarray}
  \xi = \frac{1}{2}(r^2 + r'^2 + |\textbf{r} - \textbf{r}'||\textbf{r} + \textbf{r}'|), \label{eqn:xi} \\
  \eta = \frac{1}{2}(r^2 + r'^2 - |\textbf{r} - \textbf{r}'||\textbf{r} + \textbf{r}'|). \label{eqn:eta}
\end{eqnarray}

The derivatives of the confluent hypergeometric functions~$U$ and~$M$ are respectively given by~\cite{macek1998}:
\begin{eqnarray}
  \frac{\partial}{\partial \xi} U(a, b, \xi) = -a U(a + 1, b+1, \xi), \\
  \frac{\partial}{\partial \eta} M(a, b, \eta) = \frac{a}{b} M(a + 1, b+1, \eta).
\end{eqnarray}
To proceed, let us further introduce the following notation for the sake of brevity:
\begin{equation*}
  F_E^{(n)}(x) \equiv F\bigg(\frac{4n-1}{4} - \frac{E}{2}, \frac{2n+1}{2}, x \bigg),
\end{equation*}
where $F$ denotes the confluent hypergeometric function~$U$ or~$M$, parameter $n$ is an integer and $x$ denotes~$\xi$ or~$\eta$ defined in Eqs.~(\ref{eqn:xi})
and~(\ref{eqn:eta}), respectively.

Substituting the derivatives into Eq.~(\ref{eqn:Greens1}), we obtain the following expression for the Green's function:
\begin{widetext}
  \begin{flalign}
    \begin{split}
      G(\textbf{r}, \textbf{r}') &= \exp{\bigg(-\frac{\xi + \eta}{2}\bigg)}\bigg\{\Lambda(1, E) U_E^{(1)}(\xi)M_E^{(1)}(\eta) + \frac{2 \xi \eta}{\xi -
        \eta}\Lambda(1, E)\bigg(\frac{3}{4}-\frac{E}{2}\bigg)\bigg(\frac{2}{3} U_E^{(1)}(\xi)M_E^{(2)}(\eta)+ U_E^{(2)}(\xi)M_E^{(1)}(\eta)\bigg) \\ &+
      \mathrm{sign}(\textbf{r}\cdot\textbf{r}')\Lambda(1, E + 1)\frac{2\sqrt{\xi\eta}}{\xi-\eta}\bigg(\frac{3}{4}-\frac{E}{2}\bigg)\bigg(\frac{2}{3}\eta
      U_{E+1}^{(1)}(\xi)M_{E+1}^{(2)}(\eta)+ \xi U_{E+1}^{(2)}(\xi)M_{E+1}^{(1)}(\eta) \bigg)\bigg\}.
      \label{eqn:greensfull}
    \end{split}
  \end{flalign}
\end{widetext}
Expanding Eq.~\eqref{eqn:greensfull} in the Taylor series, we obtain the following  asymptotic behavior for in the limit when $\mathbf{r}'$ approaches  $\mathbf{r}$:
\begin{equation}
G(\textbf{r}, \textbf{r}')  \xrightarrow[]{\Delta r\rightarrow 0} g^0(R) + \frac{g^1(R)}{\Delta r},  \label{eqn:greenExp}
\end{equation}
where the distance between the points $\mathbf{r}$ and $\mathbf{r}'$ is denoted by $\Delta r = |\textbf{r}-\textbf{r}'|$, the mean position is given by $R = |\textbf{r} +
\textbf{r}'|/2$, and the function $g^0$ and $g^1$ are respectively given by:
\begin{widetext}
  \begin{subequations}
    \begin{eqnarray}
      g^0(R) &=&
      -\frac{1}{4}\Lambda(1, E)\exp(-R^2) \bigg\{ 4R M_{E}^{(1)}(R^2 ) U_{E}^{(1)}(R^2 )-\frac{1}{3} (2E-3)R^5 \bigg( 2(2E-7)M_E^{(3)}(R^2 ) U_E^{(3)}(R^2 )  \nonumber\\
      &&
      + 10(2E-3)M_E^{(2)}(R^2 ) U_E^{(2)}(R^2 ) \bigg) + 15(2E-7)M_E^{(1)}(R^2 ) U_E^{(3)}(R^2 ) \bigg\} \nonumber\\
      &&
      -\frac{1}{60} \mathrm{sign}(\textbf{r}\cdot\textbf{r}')(2E-1)\Lambda(1, E+1)\exp(-R^2)R^2 \bigg\{ (-5+2E) R^2 M_{E+1}^{(3)}(R^2 ) U_{E+1}^{(1)}(R^2 )\nonumber\\
      &&
      - 10 M_{E+1}^{(2)}(R^2 ) \bigg( U_{E+1}^{(1)}(R^2 ) + (2-4E)R^2 U_{E+1}^{(2)}(R^2 )\bigg) \nonumber\\
      &&
      +15M_{E+1}^{(1)}(R^2 )\bigg( U_{E+1}^{(2)}(R^2 ) +  (2E-5)R^2 U_{E+1}^{(3)}(R^2 )\bigg) \bigg\}, \\
      g^1(R) &=&
      - \frac{1}{12}\Lambda(1,E)( 2E-3)\exp(-R^2)  \bigg( 2 M_E^{(2)}(R^2 ) U_E^{(1)}(R^2 ) +3 M_E^{(1)}(R^2 ) U_E^{(2)}(R^2 )\bigg)   \nonumber\\
      &&
      - \mathrm{sign}(\textbf{r}\cdot\textbf{r}')\frac{1}{12}\Lambda(1, E+1) (2E-1) e^{-R^2}R^3 \bigg( 2 M_{E+1}^{(2)}(R^2 ) U_{E+1}^{(1)}(R^2 ) +
      3 M_{E+1}^{(1)}(R^2 ) U_{E+1}^{(2)}(R^2 )\bigg). \quad
    \end{eqnarray}
  \end{subequations}
\end{widetext}


\providecommand{\noopsort}[1]{}\providecommand{\singleletter}[1]{#1}%

\end{document}